# Electron backscattering for signal enhancement in a thin-film CdTe radiation detector

## Fatemeh Akbari, Diana Shvydka*

Department of Radiation Oncology, University of Toledo Health Science Campus, Toledo, OH, USA

* Corresponding author email: Diana.Shvydka@utoledo.edu

## Abstract

**Background:** Thin-film Cadmium Telluride (CdTe) offers high average electron density, direct detection configuration, and excellent radiation hardness, making it an attractive material for radiation detectors. Although a very thin detector provides capabilities to conduct high resolution measurements in high energy radiation fields, it is limited by a low signal, often boosted with a front metal converter enhancing x-ray absorption. An extension of this approach can be explored through investigation of electron backscattering phenomenon, known to be highly dependent on the material atomic number Z. Hence with proper design metal back electrode can be utilized for signal enhancement. Electron backscattering coefficients from a multilayer structure provide useful information on expected signal boosting in a thin-film device.

**Purpose:** We investigated the possibility of augmenting the fluence of electrons traversing CdTe thin film and thus increasing the detected signal pursuing two venues: (1) adding a high-Z metal layer to the back of the detector surface, and (2) adding a top low-Z material to the detector layer to return its backscattered electrons. Copper (Cu) and lead (Pb) layers of varying thickness were investigated as potential metal back-reflectors, while polymethyl methacrylate (PMMA) water phantom material was tried as the top cover in multilayer detector structures.

**Methods:** The Monte Carlo (MC) radiation transport package MCNP5 was first used to model a basic multilayer structure, including a CdTe sensitive volume surrounded by PMMA, under a Varian linac's 6MV photon beam. It was then modified by the addition of Cu and Pb metal back-reflectors to analyze the extent of the signal enhancement and associated changes in secondary electron fluence spectra. Backscattering coefficients were then calculated using EGSnrc MC user-code for a set of monoenergetic electron sources. Analytical functions were established to represent the best-fitting curves to the simulation data. Finally, electron backscattering data were related to signal enhancement in the CdTe sensitive layer based on a semi-quantitative approach.

**Results:** We studied multilayer detector structures, decoupling the effects of PMMA and the back-reflector metals on detector backscattering properties. It was found that adding a metal film below the sensitive volume of a detector increases the fraction of reflected electrons, especially in the low energy range. The thickness dependence of backscattering coefficients from thin films exhibits saturations at values significantly exceeding the electron ranges. That effect was related to the large-angle electron scattering. A detailed simulation of energy deposition revealed that the modified structures using Cu and Pb increased energy deposition by ~10% and 75%, respectively. We have also established a linear dependence between the energy deposition in the semiconductor layer and the fluence of backscattered electrons in the corresponding multilayer structure. The low-Z top layer in practically implemental thicknesses of tens of microns has a positive effect due to partial electron reflection back to the semiconductor layer.

**Conclusions:** Signal enhancement in a thin-film CdTe radiation detector could be achieved using electron backscattering from metal reflectors. The methodology explored here warrants further studies to quantify achievable signal enhancement for various thin-film and other small sensitive volume detectors.

# 1- INTRODUCTION

Semiconductor materials offer the convenience of a direct signal detection under high energy photon beams, utilized in diagnostic radiology and radiation therapy. They are often employed in both dosimetry and imaging applications, with the latter option requiring large area devices, much easier implemented with thin films. For historical reasons, the choice of materials is limited to amorphous silicon (a-Si), while manufacturing technology for other better suited materials, such as cadmium telluride (CdTe), have been firmly established in photovoltaic applications.[2] Compared to a-Si, thin-film CdTe offers superior efficiency with higher average electron density and direct detection configuration, as well as outstanding radiation hardness.[3,4] With a typical device thickness below 1 mm, low absorption efficiency in high energy photon beams would benefit from signal boosting. One established approach is to use a metal "converter" plate above the thin film[3], where the plate increases photon interactions and thus an influx of secondary electrons reaching CdTe sensitive volume. Here we propose an approach to enhance the detectable signal with use of metal back-reflectors, relying on a known phenomenon of electron backscattering.

When electron beam is directed onto the surface of a solid target, some electrons emerge from the incident surface due to elastic and inelastic collisions with the atomic electrons and the nuclei of the target medium. This process is often characterized in terms of the backscattering coefficient ($\eta$), defined as the ratio of the backscattered electrons to the total number of the incident electrons. The electron backscattering is of great interest in numerous applications such as scanning electron microscopy (SEM), electron microlithography, Auger electron spectroscopy, and studies of radiation damage. The phenomenon is also important in medical physics for the purpose of accurate assessment of dose deposited around inhomogeneities where backscattering alters the spatial distribution pattern.

Electron backscattering from bulk specimens, thin films and their combinations has been a subject of many studies. Several attempts have been made to describe the electron backscattering process by a simple theory.[5-11] In some applications, single scattering processes are insufficient to characterize electron backscattering. Existing multiple scattering theories, on the other hand, usually only describe a few limiting conditions or are difficult to assess. Most of the information comes from the experimental data, covering a wide range of incident angles and energies, target samples, and experimental geometries.[12-16] Most of these studies have been conducted for incident electron energies less than 140 keV[7,10,14,16-18], a few at high energies (E>1 MeV)[12,13,15,17,19]. Little data exists in the intermediate range (0.1 MeV< E <1 MeV)[19-21], which can be important for various spectroscopy applications, as well as in measurement of radiation fields in radiotherapy applications. Furthermore, electron backscattering coefficients have been obtained with Monte Carlo (MC) simulations using different codes including MCNP, GEANT, EGSnrc, CASINO, and some others[17-19,21-34]; here the typical approach is to calculate the fraction of electrons scattered from a sample surface and returned into vacuum. The published data is usually described with empirical fitting functions, which are difficult to generalize to arbitrary targets, especially in multi-layer structures.

The main purpose of this study is to explore the potential of modifying electron spectrum and boosting signal in a semiconductor radiation dosimeter utilizing electron backscattering. We conduct MC simulations of multilayer structures based on CdTe in combination with two significantly different metal back reflectors, copper (Cu) and lead (Pb). The former option was selected due to its ubiquitous use in electronic circuits, while the latter offers high backscattering coefficient owing to its high atomic number[5]. The effect of polymethyl methacrylate (PMMA), which is a water equivalent material often used for detector encapsulation, is also investigated as a part of the detector design. Starting with a photon source modeling and establishing relevant parameters for the generated secondary electrons we investigate the basic and modified structures under a set of monoenergetic electron sources. Finally, analytical functions that represent the best-fitting curves to the simulated data were established. By de-coupling the effects of PMMA and the back-reflector metals on CdTe backscattering properties we can offer a general description of the multi-layer detector properties without the need of simulating all conceivable layer configurations. Our approach can be generalized at least semi-quantitatively to alternative semiconductors and back reflectors.

## 2- MATERIALS AND METHODS

### 2-1 Detector design

Monte Carlo simulations (MCNP5 package[35]) were conducted first to model a basic multilayer structure, including a CdTe sensitive volume under 6MV photon beam of a Varian linac[36]. The basic structure was then modified by the addition of a metal back-reflector serving to increase its detected signal. Schematic representation of the modeled detector design is shown in Figure 1, where the arrow on the right represents the direction of "backscattered" electrons, defined as any scattered or secondary electrons moving in the direction towards the primary source.

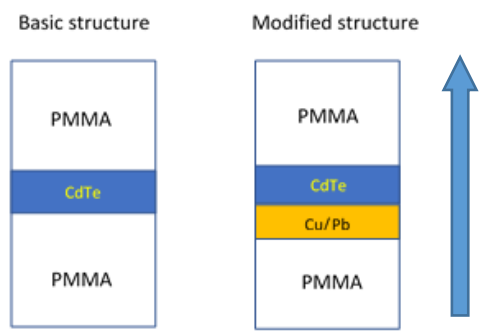

Figure 1: Schematic representation of the modeled detector design: basic structure and modified structure, including back reflector (not to scale). The arrow represents the direction of "backscattered" electrons

CdTe of 30μ and 300μ thickness was simulated in combination with copper and lead, with surrounding phantom-equivalent PMMA layer. For 6MV photons, deposited energy "builds up" to an equilibrium value over a thickness range of ~1.5cm. Thus, in addition to serving as a detector encapsulating layer, the PMMA layer in our structure also performed that function. Photon interactions set secondary electrons in motion, and these electrons deposit their energy (dose) within the detector sensitive volume. Photon simulation was used to look at the generated secondary electron energy spectra at the interfaces between PMMA, semiconductor, and metal reflectors. It informed the next step of Monte Carlo simulations under monoenergetic electron sources, defining the range of energy for backscattered electron fluence calculations.

The saturation thickness for which backscattered fraction becomes maximum (it is about the half of electron range for a specific energy) was used as the optimum thickness of the metal back-reflector layer. Range of electrons of different energies in materials of interest and properties of the materials are provided in Table 1. In all the simulations, we set cutoff energies to 10 keV for electrons and photons, with coherent, photonuclear, and Doppler interactions turned off, but Bremsstrahlung included. The number of particles crossing a surface was calculated using F1 tally (surface current) and cosine binning to distinguish between forward and backscattered particles. *F8 energy deposition tally was used to acquire relative signal in studied structures; the tally value was divided by the mass of the tally cell to obtain the energy deposition per unit mass of the detector sensitive volume. $3\times10^8$ photon histories were followed to achieve statistical errors less than 1%.

Table 1. Summary of modeled material properties: density, mean atomic number (Z), and range of electrons (in µm) for different energies

| Material | Density (g/cm³) | Atomic number | E (keV) | | | | | | | |
|---|---|---|---|---|---|---|---|---|---|---|
| | | | 20 | 50 | 100 | 150 | 200 | 300 | 500 | 700 |
| PMMA | 1.18 | 6 | 7 | 37 | 124 | 243 | 387 | 727 | 1525 | 2403 |
| CdTe | 6.20 | 52 | 3 | 15 | 46 | 88 | 138 | 254 | 519 | 802 |
| Cu | 8.96 | 29 | 2 | 8 | 25 | 48 | 76 | 140 | 292 | 456 |
| Pb | 11.35 | 82 | 2 | 9 | 27 | 52 | 81 | 147 | 296 | 454 |

## 2-2 Electron backscattering

Ali and Rogers[22] developed and made available a customized EGSnrc MC user-code for backscattering coefficient calculations. Since the code conveniently provides backscattering coefficient value for a customizable structure, it was utilized in this work to obtain $\eta$. The code simulates an electron or positron beam incident on a sample (4<Z<92), which may be a thin film, a stack of films, or a bulk target considered infinite if its thickness is larger than the range of incident charged particles in the sample. The code validation against experimental data from 20 separate published experiments encompassing 35 distinct elements, electron and positron backscattering, normal and oblique incidence, confirmed the simulation results to be within 4%[22]. Although accurate charged particle backscattering simulation could be challenging for both the EGSnrc and MCNP codes that use the condensed history algorithm, it has been shown that implementation of this technique in the EGSnrc code yields results that are step-size independent, and agree with single scattering (condensed) calculations within less than 0.1% for low energies[37]. In this work η was calculated for CdTe with added Pb or Cu reflector layer of varying thickness, 20 to 200 microns, and a top layer of PMMA of 1 to 50 micron thickness. The model comprised of a pencil beam of monoenergetic charged particles incident normally on a thin film sample or a stack of films, surrounded by vacuum. The geometry of each simulated structure is depicted in Figure 2. Backscattered charged particles were tallied as they cross from the sample medium back to the vacuum. The values of electron and photon transport cut off used in the simulations are 512 and 1 keV, respectively. For electrons, this value corresponds to a kinetic energy of 1 keV. A total of $5\times10^4$ electron histories were run; the maximum uncertainty in Monte Carlo calculations of $\eta$ values and energy spectra were 1% or lower.

The first step in our electron transport modeling involved verification of the appropriateness of MC packages used for thin film simulations, in view of their use of condensed history algorithm. In MCNP this approximation is defined by a parameter called DRANGE, which is the size of an energy step in g/cm$^2$. It is further divided into a number of supsteps empirically determined to be in the range of 2 to 15, depending only on the average atomic number of the material.[35] A rule of thumb for the appropriate number is to ensure that electrons make at least ten substeps in any material relevant to the transport problem, thus the size of a substep should be compared to the smallest material dimension. Our problem geometry did comply with this rule; additionally, we verified that setting this parameter to its maximum value of 15 had virtually no effect on the results of our simulations.

Backscattering coefficient was calculated as the ratio between the total number of electrons reflected from the first non-vacuum layer and those entering the top layer. Electron backscattering coefficients obtained by the EGSnrc user-code were also verified against MCNP simulations, utilized to acquire the data not available from the EGSnrc user code. The change in electron backscatter and forward fluence was also investigated.

In order to obtain the value of $\eta$ for the arbitrary target thickness combined with a metal back-reflector or surrounded with a PMMA layer, and for the incident kinetic energy E of the electrons, it would be convenient to derive an equation which well reproduces the most probable values given by the existing data. Therefore, the best fit for energy-dependent backscattering coefficients, characterized with the least R-square value, was also obtained for all structures for prognostic purposes.

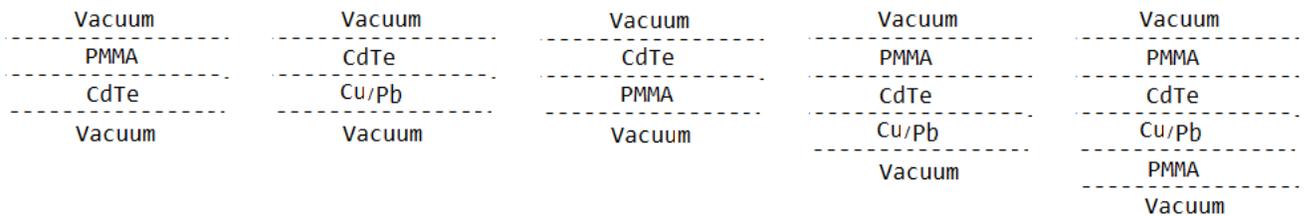

Figure 2: Simulations geometries used to calculate $\eta$ at the first interface (top dashed line)

## 2-3 Estimate of signal enhancement based on the electron backscattering

Backscattering coefficients obtained for modified structures with a back reflector were used for a semi-qualitative estimate of signal enhancement in the CdTe sensitive layer. Specifically, increase in the reflection coefficient evaluated at the top of CdTe is indicative of additional traversing of the detector layer by electrons reflected from the interface with metal. These electrons must have larger range than CdTe film thickness and produce signal enhancement in CdTe. As this effect is observed at higher electron energies, the relative fractions of those electrons could be used for the signal increase estimate.

## 3- RESULTS

This section contains results for backscattering electron spectra under a 6MV photon beam, electron backscattering coefficients, modeled with monoenergetic electron sources, from metal back-reflectors, Cu or Pb, of various thicknesses, and backscattering data from combinations of CdTe with PMMA/Cu/Pb. The Figures and Tables show energies and thicknesses in units of keV and microns, respectively. In legends and other labels a number following each material represents its thickness.

### 3-1 Secondary electrons generated under 6MV x-ray source

Detector structures of Figure 1 were first modeled with the 6MV photon source to evaluate the overall effect of metal back-reflector on production and scattering of secondary electrons traversing CdTe sensitive volume. Increase in the secondary electron fluence leads directly to the detector signal enhancement. Example of the change in the secondary electron fluence at the top and bottom surfaces of the sensitive volume of the basic structure (30 μm thick CdTe only) upon addition of the metal layer in the modified designs is presented in Figure 3. Similar results were obtained for structures with 300 μm thick CdTe layer.

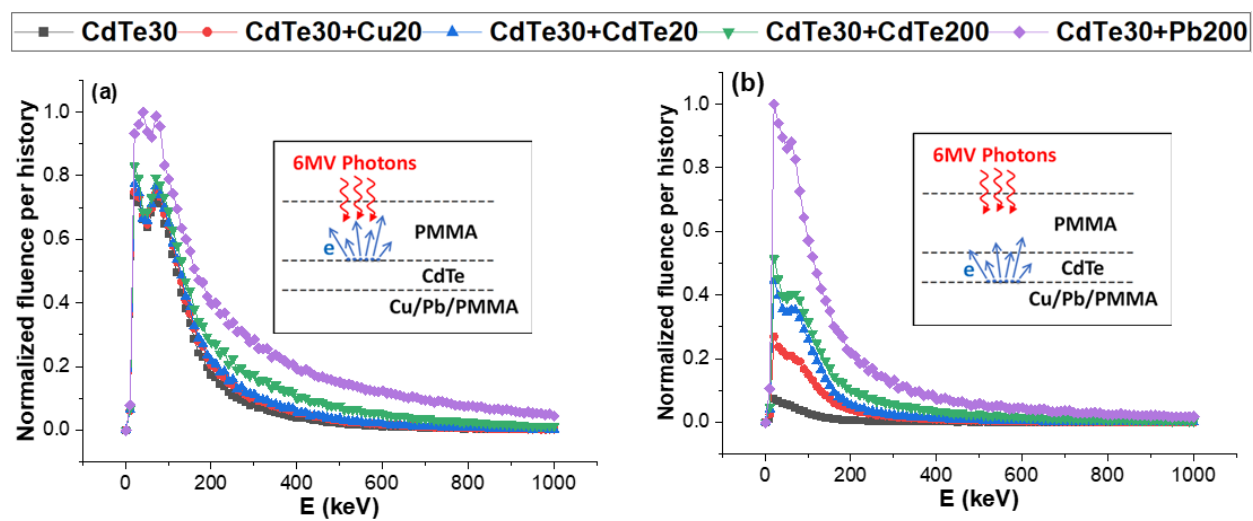

Figure 3: Normalized fluence per history for the secondary electrons, generated under 6MV photon beam and moving in the directions opposite to those of the source photons. a) The spectra obtained at the top surface of the 30μ CdTe detector b) the fluence spectra at the bottom surface of the 30μ CdTe detector sensitive volume, and the modified structures using 20μm Cu, 200μm Pb, and additional CdTe layers of the same thicknesses. The inserts show sketches of simulation geometry.

Based on spectra shown here, the secondary electrons moving in the directions opposite to those of the source photons from the bottom surface of the detector can be drastically affected by presence of the backscattering layer. Using Pb and Cu, the total fluence of electrons scattered back in CdTe with a metal back-reflector and CdTe alone were found to be increased by factors of 24 and 5, respectively. This suggests that the detector signal can be adjusted in a multilayer structure consisting of a sensitive volume and a metal back-reflector layer. For direct comparison structures with equivalent thickness of CdTe were also simulated. While the backscattered electron fluence with additional Cu layer (CdTe30+Cu20) is slightly lower than that with the additional CdTe layer of the same thickness (CdTe30+CdTe20), the advantage of using metal instead of semiconductor is that the metal layer can serve as an electrode, thus allowing for lower bias applied across a thinner sensitive layer to achieve very similar signal.

The average energy of the electrons reflected from the bottom of CdTe was found to increase with addition of a backscattering layer, especially for Pb, having the highest atomic number. Table 2 summarizes main characteristic parameters of fluence spectra in Figure 3, such as "backscattering" electrons fluence ratio (area under each graph in Figure 3, where CdTe30 structure is taken as unity) and deposited dose in CdTe volume. A dose ratio comparison between different configurations and CdTe only is also provided as an illustration of practically achievable signal enhancement.

The energy of the majority of generated secondary electrons is limited, becoming negligible above 1 MeV. As a result, we only focus on electrons up to 500 keV for basic CdTe and CdTe/Cu structures, extending it above 1 MeV only for CdTe/Pb configuration.

Table 2. Summary of main characteristic parameters for electron fluence spectra of Figure 3 (all data obtained under 6MV x-ray source)

| Structure | Fluence ratio | | Dose (MeV/g) | Dose ratio |
|---|---|---|---|---|
| | Top surface of CdTe | Bottom surface of CdTe | | |
| CdTe30 | 1 | 1 | 8.50 | 1 |
| CdTe30+Cu20 | 1.11 | 5.04 | 9.20 | 1.08 |
| CdTe30+CdTe20 | 1.15 | 7.95 | 9.81 | 1.15 |
| CdTe30+CdTe200 | 1.45 | 11.54 | 10.82 | 1.27 |
| CdTe30+Pb200 | 2.17 | 24.42 | 14.80 | 1.74 |

Since the detector signal enhancement is primarily due to secondary electron backscattering from the back CdTe/metal layer, we concentrate on electron transport properties in the basic and modified structures in what follows.

## 3-2 Electron backscattering coefficient

### 3-2-1 Interactions of monoenergetic electrons in CdTe

To illustrate the electron interaction process we examined the change in backscattered and forward electron fluences correspondingly at the top and bottom surfaces of CdTe layer under monoenergetic electron sources. MCNP simulations were used to calculate the energy spectra shown in Figure 4 for a film of 30μm thick CdTe. The monoenergetic electron sources, 100 to 500keV, are labeled in the graph. A good agreement for backscattering calculations using MCNP and EGSnrc for a 300keV electron source is presented as an insert in portion (a) of this figure. For a better presentation in part (b) of the figure, the forward electron fluence utilizing a 100keV electron source has been amplified by a factor of 100.

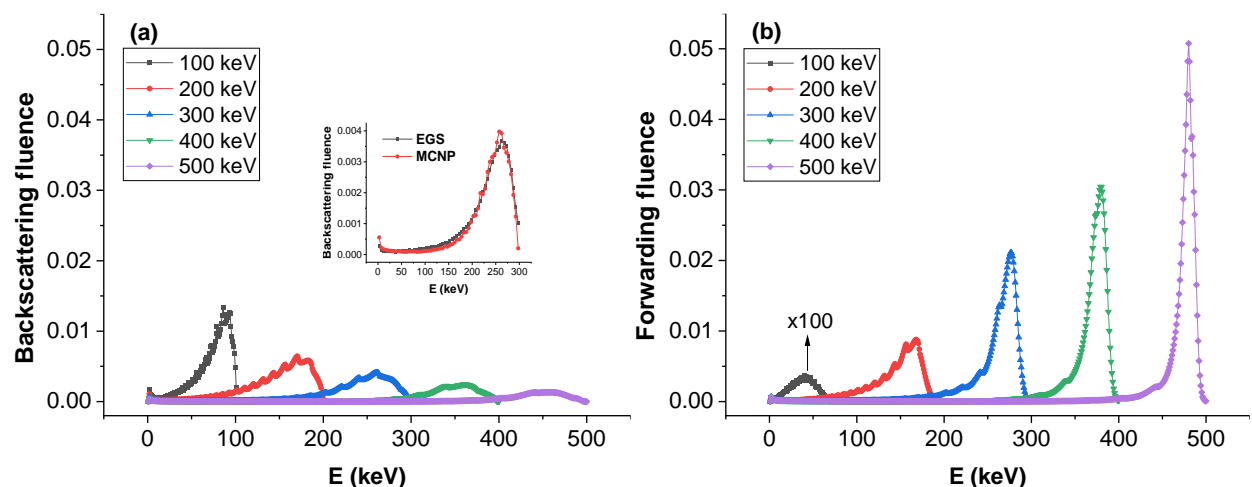

Figure 4: (a) Backscatter fluence spectra from top surface of CdTe30, insert shows a side-by-side comparison of the backscatter fluence calculated by EGSnrc and MCNP codes for incident electrons of 300 keV. (b) Forward fluence at the bottom of CdTe30. The data from a 100 keV source is multiplied by 100.

The same number of source particles was used in simulations with different energies. Significant energy dependence of the backscattering process illustrated in Figure 4 is further investigated through changes in backscattering coefficients in multilayer structures presented below.

### 3-2-2 Saturation thickness of metal back-reflector

Electron backscattering from thin films is known to depend on the material thickness, reaching a bulk specimen value when the layer thickness becomes about twice the electron range in the material. Figure 5 shows electron backscattering coefficient at Cu (Pb) and vacuum interface for varying film thickness and an energy range of interest, established based on the photon simulation results. The intensity of backscattered electrons increases with the target thickness, saturating beyond a certain amount known as the saturation thickness. The figure also illustrates differences between maximum backscattering from metals with a medium and high atomic numbers, Cu and Pb. While these values were obtained with EGSnrc user code, a similar simulation for a subset of configurations using MCNP revealed good agreement, within 2%, in backscattering coefficient.

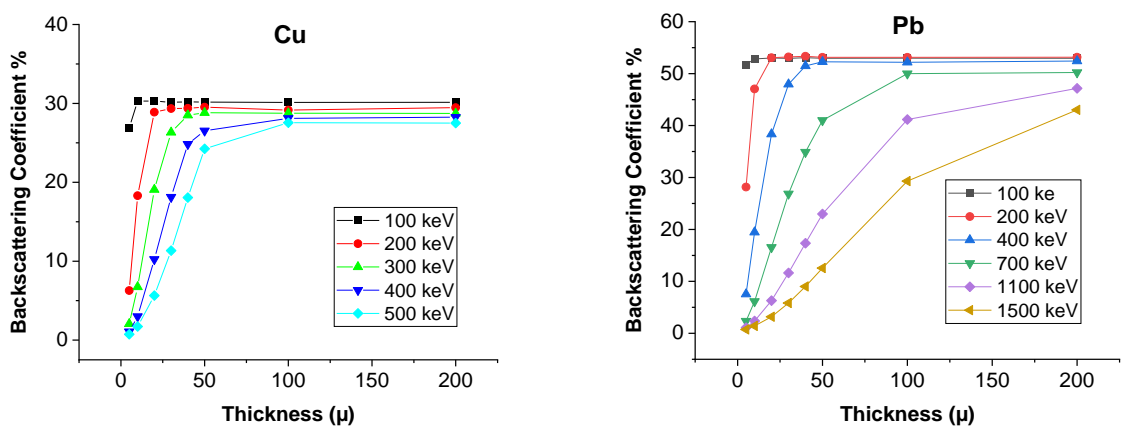

Figure 5: Backscattering coefficient versus thickness of metal back-reflectors of Cu and Pb

### 3-2-3 η and electron source energy

While the main focus of this investigation is the effect of the back-reflector, we start with evaluation of the omnipresent phantom-equivalent PMMA layer on top of CdTe. Figure 6 shows the electron backscattering coefficient as the function of the electron source energy for modified detector structures involving combinations of 30 and 300-micron CdTe with PMMA (on top), and Cu and Pb (at the bottom) layers of varying thicknesses. Numbers in the legends represent the thickness in microns, e.g., CdTe30 stands for CdTe only configuration, having thickness of 30μ, or PMMA1+CdTe300 represents a structure with 1μ thick PMMA over 300μ thick CdTe. Symbols in Figure 6 show simulation results, the solid lines represent the best fits with the BiHill function[1], which was utilized for all modeled structures.

Figure 6(a,b) show the η-values obtained for structures of PMMA of various thicknesses on top of 30 and 300μm thick CdTe. The addition of the PMMA layer, required to account for realistic encapsulated detector configurations, results in a reduction in η, more significant for the thinner CdTe. Figure 6(c) illustrates the electron backscattering coefficient for CdTe30+Cu structure at various energies. Figure 6(d), on the other hand, depicts a pattern that is consistent across all combinations indicating that the thick CdTe layer of 300μ dominates the backscattering process for the relevant energy range. Figure 6(e) shows backscattering from CdTe and a high atomic number metal back-reflector of Pb over a wider range of energies. Figure 6(f) shows a pattern overall very similar to combination effects of thick CdTe and copper.

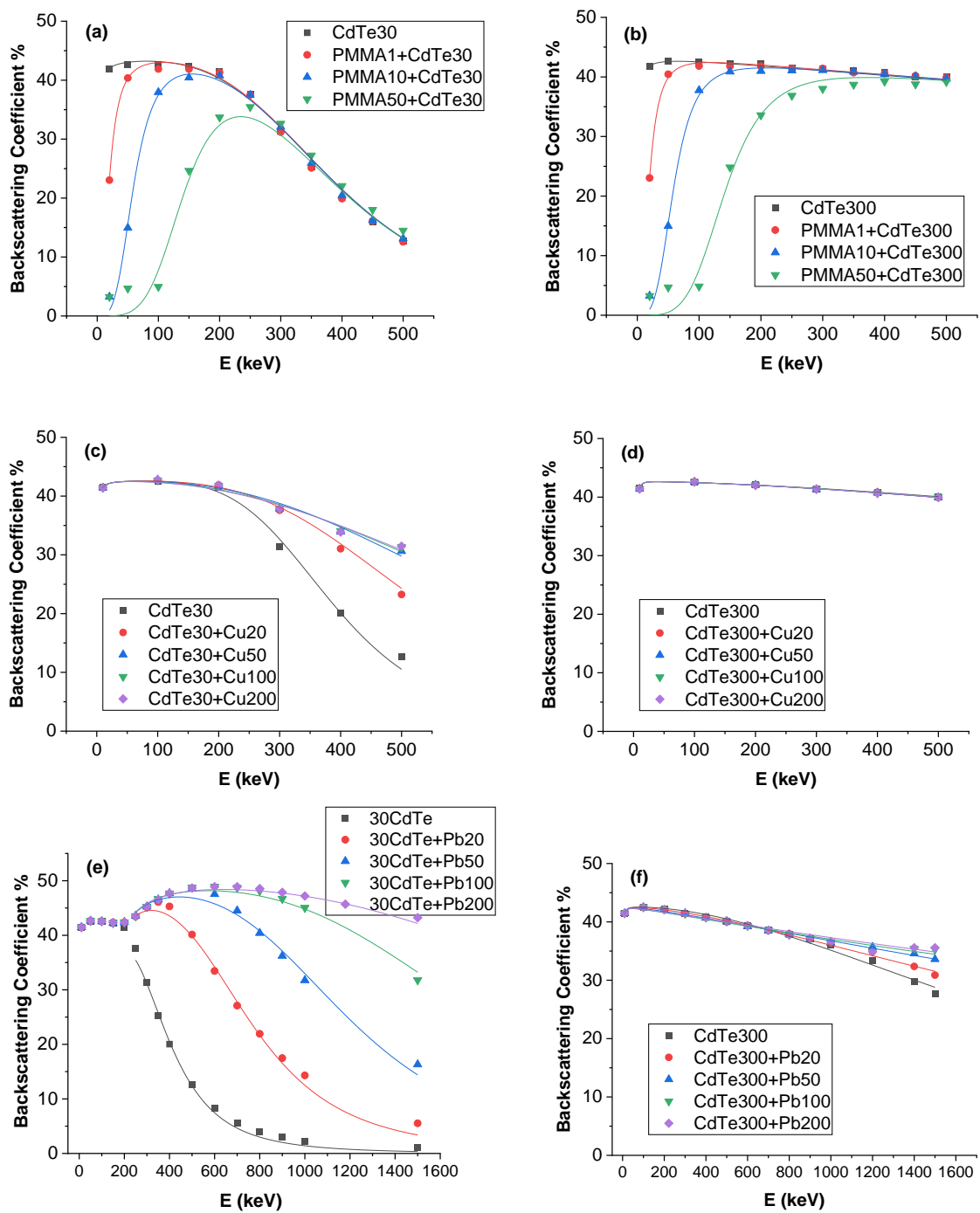

Figure 6: Electron backscattering coefficient (η) dependance on the electron source energy for various combinations of PMMA on top of CdTe, or Cu/Pb at the bottom of CdTe. Numbers in the legends represent thickness in microns, e.g., CdTe30 stands for CdTe only configuration, having thickness of 30μ. Symbols show simulation results, and the solid lines represent the best fits with BiHill function (see Appendix). Left: with 30μ CdTe, right: using 300μ CdTe.

The maximum electron backscattering coefficient was found to be 44%, 42.7%, and 50% for 30µ CdTe combinations with PMMA, Cu, and Pb, respectively. The position of the peak value shifts toward higher energies as the PMMA or metal thickness increases. When combined with PMMA, 300µ CdTe has a maximum backscattering of 44%, while Cu and Pb both have a maximum backscattering of 42.7%. The modified structures of CdTe and underlying Cu or Pb metal back-reflectors produce different patterns of overlapped and splitted curves in the graphs.

The simulation results were fitted with BiHill function using Origin 9 software.[1] Basic description of this function properties and all fitting parameter values can be found in the Appendix. While the function has 5 fitting parameters, for most of the evaluated structures only 2 or 3 parameters were varied, the rest, describing the part of the graph with all curves collapsing on top of each other, were fixed. For all fits, R-square value of 0.98 or higher was achieved.

Another representation of the overall effect of combination of materials with significantly different atomic numbers is presented in Figure 7, using PMMA and CdTe for illustration. As evident from the figure, electron backscattering coefficients for the bi-layer structures are limited between η-values of pure PMMA (the lowest curve) and pure CdTe (the highest curve). Bi-layer values start from being equal to those of pure PMMA but then increases as the electron energy increases to achieve the values of CdTe, at energies depending on the PMMA thickness. Electrons of low energies cannot reach the CdTe layer and η reflects backscattering from top layer only. For example, for PMMA of 10µ thickness, according to Table 1, maximum range of low energy electrons (below 20 keV) is 7µ which is not large enough for a backscattered electron to exit PMMA layer. As the energy increases above 50 keV, the range increases to at least four times greater than the PMMA thickness. Thus, electrons pass through to the CdTe layer and η increases toward the values of CdTe target. The trend of decreasing the backscattering coefficient for the low energy electrons is independent of the CdTe thickness. The shape of the graphs in Figure 6(a, b) can be explained in a similar way.

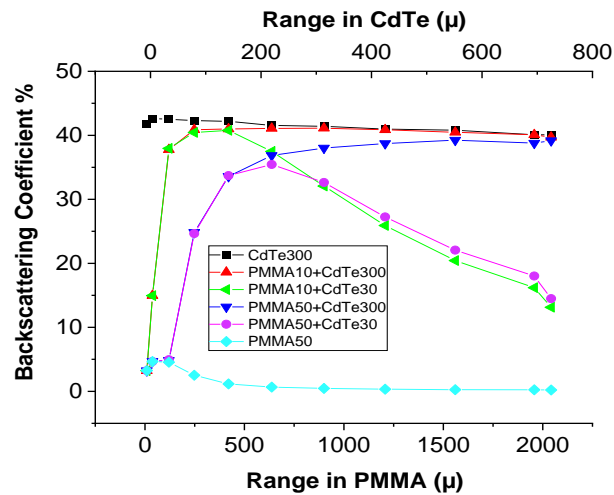

Figure 7: Backscattering coefficient versus electron range in PMMA and in CdTe

It is worth noting that the existence of a PMMA layer on the opposite side of the sensitive volume, i.e. at the bottom of CdTe, is minimal. As illustrated in Figure 8, backscattering coefficients follow the declining with energy, a trend typical of the CdTe30 backscattering coefficient with the same pattern for CdTe300.

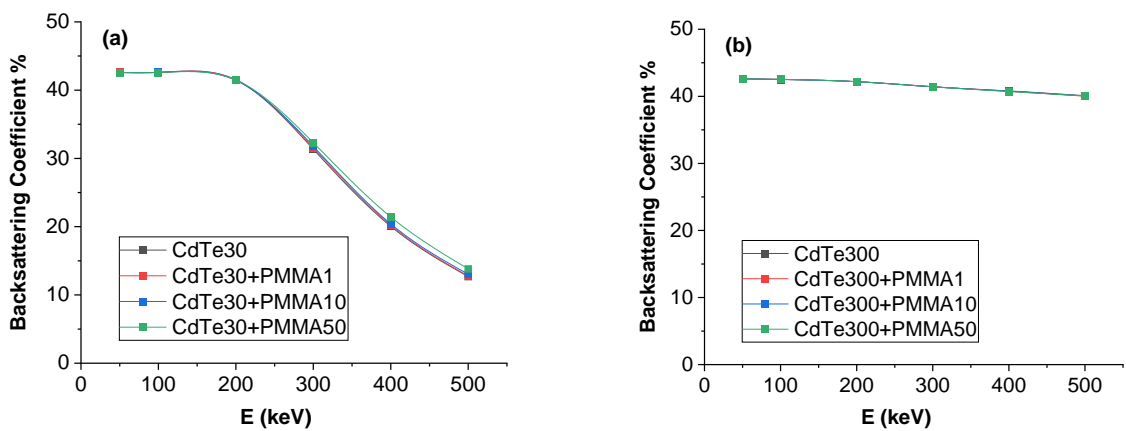

Figure 8: Electron backscattering coefficient from CdTe with PMMA underlying layer

Finally, backscattering coefficients for the whole multilayer structure consisting of CdTe and metal surrounded by PMMA on both sides were derived in order to extrapolate the results to the real model of a detector (modified structure in Figure 1). As presented in Figure 9, the general trends for selected combinations of four layers, are equivalent to those in graphs of Figure 6 and 8, where only two layers were modeled. Thus the behavior of the backscattering coefficient under monoenergetic electron sources can be presented in a de-coupled fashion, representative of the overall multi-layer structure.

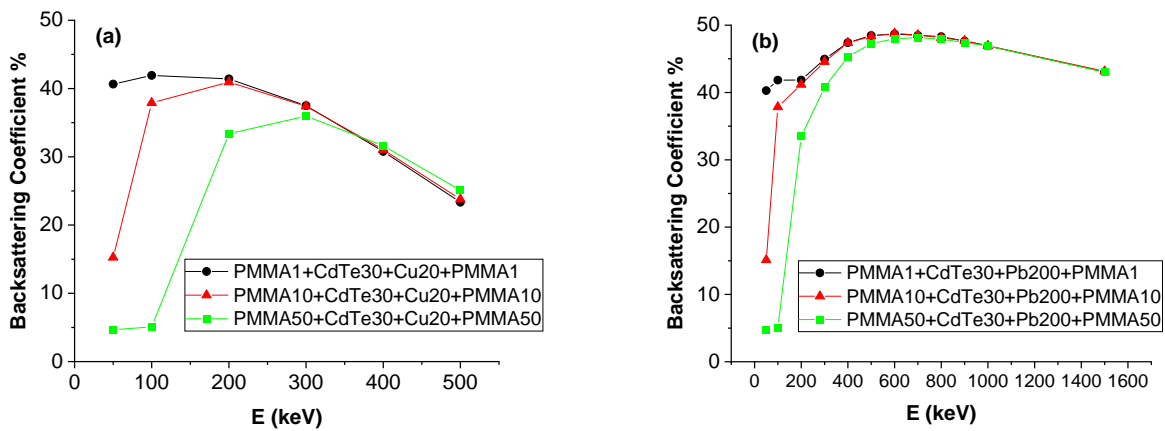

Figure 9: Backscattering coefficient from a modified multilayer structure shown in Figure 2, including four layers of PMMA+CdTe30+Cu/Pb+PMMA structure. (a) and (b) show models using Cu and Pb, respectively

### 3-2-4 Backscattering coefficient data and signal enhancement

Here we will attempt to unify observations under 6MV photon source in realistic detector structures (Figure 3) and the reflection coefficient results obtained under electron sources. Generally speaking, the fluence of electrons moving towards the source (broadly named "backscattered") in Figure 3 from either top or bottom surface of CdTe may not be a good indicator of the dose deposition within CdTe layer. The following two obvious factors complicate the relation between these parameters. 1) Due to the angular dependence, a substantial fraction of scattered or secondary electrons will travel long distances in the lateral directions, some not even emerging from the CdTe layer. While not accounted by the electron fluence they will nevertheless contribute to the energy deposition. This effect is illustrated more in detail in the next section. 2) On the other hand, fast electrons traverse CdTe film without much interaction; their fraction is relatively low, decreasing as $1/E^2$ with energy.[38] Such electrons will contribute more to the fluence count than to the dose, they emerge from the thin film retaining most of their kinetic energy.

A detailed simulation of the energy deposition under the photon source shows indeed increase in the modified structures compared to the basic one of a single CdTe layer. Further analyzing the summary of simulation parameters presented in Table 2, we found a very strong correlation between the "backscattered" electron fluence ratios of Figure 3 and dose deposition. The corresponding plots are shown in Figure 10(a) and (b) for the fluence ratios for the top and bottom CdTe surfaces. In view of the above factors, our established linear dependence between the fluences and doses appears rather nontrivial. It can perhaps be attributed to the fortunate mutual compensation of the above mentioned opposite trends in the simulated data where the fluences are integrated over the entire electron spectra. More discussion is provided in Section 4 below.

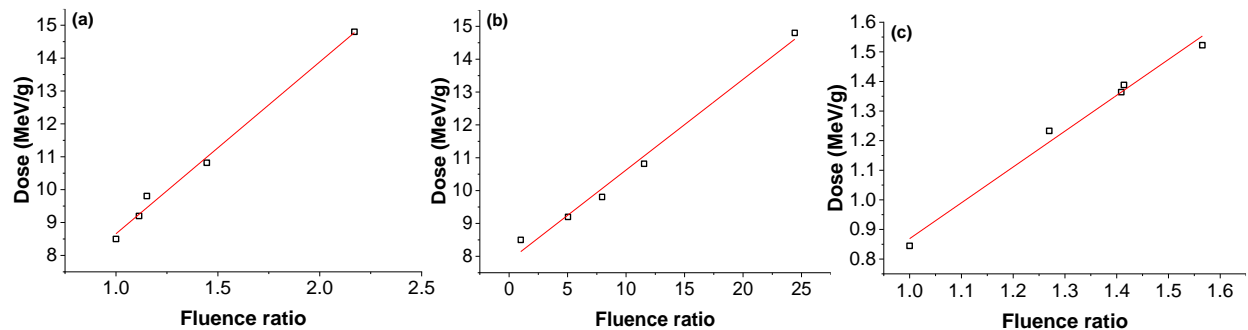

Figure 10. Relationship between the secondary electron fluence ratios and dose deposition in CdTe layer obtained under 6MV photon source (based on Figure 3 and Table 2) from the top (a) and the bottom (b) of CdTe layer. The data for 300keV electron source (c) is shown for comparison. Solid lines represent linear fits.

These considerations are conceptually similar to those obtained for the backscattering coefficient, showing increase in the $\eta$ value for modified structures. As evident from Figure 6, the metal reflector contributes significantly to the backscattering coefficients of thinner, 30μm, CdTe structures at higher energies. Above 300keV the $\eta$ value increases by about 25% for CdTe/Cu, and by about 50% for CdTe/Pb, indicating proportional increase in high energy backscattered electron fluence. At this energy the electron range is much larger than the semiconductor thickness, allowing electrons reflected at various angles from the interface with metal (bottom of CdTe) traverse it another time, thus enhancing the energy deposition in CdTe. Plotting again a relationship between backscattered electron fluence ratio and the dose deposited in CdTe for the same configurations as considered in Figure 3, except with CdTe, or bi-layer structures surrounded by vacuum, as for all the $\eta$ value simulations, we also find linear dependence shown in Figure 10(c). The electron source energy here was 300keV, the dose deposited is much lower than in structures surrounded by the equilibrium layers of PMMA in photon simulations.

## 4- DISCUSSION

In a detector under a megavoltage photon beam, the photons first transfer their energy to secondary electrons. Set in motion, the latter interact directly with the charged particles within the sensitive volume of the detector, creating electron-hole pairs collected at the detector terminals. Following this general process of the detector signal generation, we first investigated the effect of modifying the basic detector structure (Figure 1) with metal back reflector. This resulted in the enhanced production and return of secondary electrons into thin-film CdTe, and a proportional increase in energy deposition, conducive of the signal increase in a real device. Obtaining the spectral distributions for generated under 6MV beam secondary electrons, returned into CdTe allowed us to simplify the study: instead of a detailed simulation of the energy deposition in CdTe for all structures of potential interest, we can evaluate the behavior of the backscattering coefficients under a limited set of monoenergetic electron sources, utilizing available user code.

The materials investigated in this work represent those with very low, medium, and high atomic numbers Z, corresponding to PMMA, Cu/CdTe, and Pb. The high Z targets exhibit stronger elastic scattering (proportional to $Z^2$), and correspondingly more significant deflections amplifying the yields of backscattered electrons. Another potentially important parameter is the excitation potential (*I-value* defined as the minimum energy required to eject an electron from an atomic shell), which increases from 74 eV for PMMA, to 823 eV for Pb; its energy loss dependence however is logarithmically weak. As a result, the approximate semi-empirical Thomson–Whiddington energy loss relation holds[38]:

$$E_0^2 - E^2 = cL \tag{Eq.1}$$

where $E_0$ is the initial energy, $E$ is the most probable energy of the electron that travelled distance $L$, and $c$ is a parameter that is proportional to Z and otherwise depends on the material parameters rather insignificantly.

The latter dependence makes its imprint on the scattering cross section,

$$\frac{d\sigma}{d\Omega} = \frac{Z^2 e^2}{4E^2} \frac{1}{(1+\cos\Theta)^2} = \frac{Z^2 e^2}{4c} \frac{1}{R-L} \frac{1}{(1+\cos\Theta)^2} \qquad \text{(Eq.2)}$$

Here R is the electron range (determined from Eq. (1) with E=0), $\Omega$ is the solid angle, and $\Theta$ is the scattering angle illustrated in Figure 11. The longer travelling distances decrease the electron energies and increase their scattering probabilities. These known results enable one to qualitatively describe the scattering features relevant here.

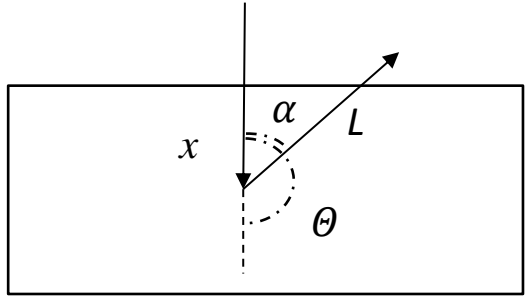

Figure 11. Sketch of a single scattering event. Scattering angle $\alpha$ defines a scattering cone in 3D geometry.

We first note that the distance $L$ traveled by the backscattered electron is significantly larger than the scatterer depth $x$. Using Eq. (2) direct averaging of the complementary scattering angle $\alpha$ in Figure 11 yields $\langle\alpha\rangle \approx 70^0$ and $L \approx 3x$. In addition, there is a significant dispersion of such angles, so some backscattered electrons travel significant distances compared to the film thickness. Their energies E and ranges substantially decrease to the extent that some do not escape the film, leading to the dose amplification.

If, on the other hand, the electron has enough energy to enter a tangent film on top of the first one, as depicted in Figure 12, then it is straightforward to see that after the second scattering, if it takes place, with the same geometry its traveled distance increases yet more, $\sim 1/cos\left(2\langle\alpha\rangle - \frac{\pi}{2}\right) \approx 5$. Traveling much longer distance, it is more likely for such an electron to slow down to the limit prohibiting its escape, which results in dose amplification.

The above remarks allow one to understand the relevance of our used film thicknesses mostly well below the ranges for electron energies summarized in Table 1. For example, the 200keV electron range of about 80µm in Pb corresponds to the Pb layer thicknesses of about 20-30µm in Figures 6(c), 6(e), according to the interpretation of Figure11, Similarly, if the top layer in Figure 12 represents CdTe, then the same energy of 200keV would correspond to the film thickness of about 30µm for CdTe in Figure 6(a), etc. In fact, all the graphs of Figure 6 can be interpreted along those lines. Our simulations presented in Figure 5 provide additional examples of electron ranges under large angles being much larger than film thickness, for example, predicting relevant Cu film thickness around 100µm for energy of 500keV that nominally corresponds to the range of 300µm, according to Table 1.

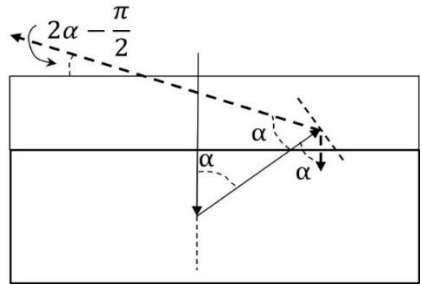

Figure 12. Sketch of the scattering geometry for a bilayer structure

Similarly, that analysis explains the role of the top PMMA layer used with some of our modeled structures. If the top layer in Figure 12 represents PMMA, which is almost "transparent" (in a sense of low scattering probability due to its low Z value) to electrons, one can see that about half of all the electrons are scattered back into the CdTe layer: hence, a useful dose amplification effect in a semiconductor layer. Note however that PMMA layer can be responsible for lower backscattering count (and its related dose deposition in CdTe; cf. Figure 10) when its thickness (by itself or in combination with other layers) becomes sufficient to completely absorb the electrons, as is illustrated in Figure 6(a). Figures 7 and 8 provide a more direct illustration of the latter statement. Note that in all the above-mentioned figures, the decay on backscattering for high energies reflects a more trivial effect of electrons penetrating through the entire structure and escaping detection.

We note that the above sketches are based on a popular 'single strong scattering' model neglecting possible diffusion propagation where electrons slow down enough to assume multiple scattering more isotropic geometry. Taking the diffusion into account will result in certain quantitative changes, retaining however the above qualitative conclusions.[4,8]

In particular, the above analysis based on the large angle scattering provides qualitative insights into the nature of low energy regions in Figures 3 and 4 as attributed to long distance traveling distance for electrons with possible multiple scattering events.

We now touch upon the derived data of Figure 10. Its presented close-to-linear positive correlations between the "backscattering" electron fluence ratios and doses deposited in CdTe layer thus far appears a strong empirical observation, with interpretation remaining challenging. At this time, we can propose some insight with a rough qualitative model illustrated in Figure 13. According to Eq. (2), the scattering cross section (translating into backscattering coefficient) increases as $1/E_0^2$ when the incident energy $E_0$ decreases. On the other hand, the stopping power is known to be inversely proportional to $E_0$. Therefore, the simplistic model of Figure 13 can indeed provide the required positive correlation between the dose and backscattering coefficient. Our attempts to prove that such a correlation must be linear (as observed in Figure 10) can be rather futile in view of the complexity of the processes involved. For example, the fluence of "backscattered" electrons will be proportional to their velocity, i. e. $\sqrt{E} \approx \sqrt{E_0 - cL}$, making the composite dependence $E_0^{-2}\sqrt{E_0 - cL}$ somewhat closer to $1/E_0$. However, subsequent integration over travel paths along with empirical modifications of the stopping power expressions[39] leave too many unknowns. Therefore, we consider the observed linear dependencies of Figure 10 as empirical, although qualitatively consistent with certain physical models.

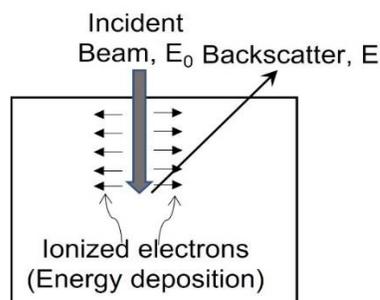

Figure 13. Qualitative model of positive correlation between backscattering and energy deposition.

Results of this work showed that adding different metal back-reflectors of varied thickness below the sensitive volume of a detector would result in a greater signal, according to "backscattered" electron fluence illustrated in Figure 3. This suggests that dose to the detector can then be adjusted in a multilayer structure consisting of a sensitive volume and a metal back-reflector layer. Figure 4 provides more information on energy deposition and backscattering from electron sources. It shows the electron spectrum of a monoenergetic electron source and how it changes when it passes through a thin-film CdTe. The spectra's peak is relatively close to the source's energy, indicating that the spectra of a monoenergetic electron source is nearly unchanged. A very low fluence of forwarding particles of 100keV source at the bottom surface of CdTe can be explained by comparing the range of these low energy electrons and CdTe thickness, according to Table 1.

## 5- CONCLUSIONS

While a very thin detector offers an extremely high resolution, desirable in some applications, it is limited by a low signal, often enhanced with a top metal converter enhancing x-ray absorption. We propose an extension of this approach with other functional layers in a detector structure. Two venues had been explored: (1) adding a high-Z metal layer on the back (instead of the typical forward) detector surface, and (2) adding a top low-Z material to return to the detector layer its backscattered electrons (in a sense, functioning as a top film in the elimination optics). We have found that both venues can provide noticeable improvements. In addition, we have attempted to relate the energy deposition in a semiconductor layer with the fluence of backscattered electrons that can sometimes be obtained more easily. Backscattering coefficient calculation was found to be in good agreement between MCNP and EGSnrc codes. However, due to its convenience, EGSnrc user code was used to collect the majority of the data.

The above work is not limited to simply juxtaposing with standard detector structures: we have explored the issues of relevant energy spectra, backscattered fluences, and their correlations with the critical detector metric of dose deposition. Overall, our created picture appears self-consistent and useful for the future detector designs.

We have observed the following:

1) Low-energy components dominate the spectra of secondary and scattered electrons generated in thin films under high-energy x-ray sources.

2) The energy spectra of monoenergetic electrons backscattered by CdTe thin films spread extensively, demonstrating wide low energy tails.

3) The thickness dependence of backscattering coefficients from thin films (studied in a wide range of parameters) exhibits saturations at values that very significantly exceed the electron ranges. That effect was shown to stem from the large-angle electron scattering.

4) The high-Z back layer (such as Pb) can increase the energy deposition by ~ 75% in the top semiconductor layer for the case of high energy x-ray sources. When such a detector is used for high energy electron source, the amplification may be much higher.

5) The low-Z top layer can be detrimental when its thickness is large enough. However, in a practically implemental thicknesses of tens of microns its effect is positive due to partial electron reflection back to the semiconductor layer. This feature can be utilized with a thin PMMA layer inserted between a top metal plate and the semiconductor layer.

6) We have established a linear dependence between the energy deposition in a detector semiconductor layer and the backscattering coefficient of its related multilayer structure.

7) We have developed a qualitative understanding of the above listed observations 1-6.

Practically speaking, we have developed an MC modeling approach for multi-layer radiation detectors. Further studies along the lines of the methodology explored here are needed to quantify achievable signal enhancement for various thin-film and other small sensitive volume detectors.

## A. Appendix: Fitting function and summary of the fitting parameters

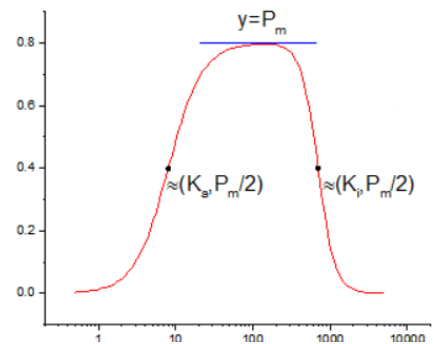

Figure 1A: Sample curve of BiHill function [1]

BiHill Function (one of the Origin library's) was used to fit energy dependance of the backscattering coefficients η. It has the form:

$$\eta = \frac{P_m}{\left[1+\left(\frac{K_a}{E}\right)^{H_a}\right]\left[1+\left(\frac{K_i}{E}\right)^{-H_i}\right]} \quad \text{(Eq.1A)}$$

where parameters are defined as $P_m$ = the value of the maximum, and $K_a$, $K_i$, $H_a$, $H_i$ are fitting parameters. We have observed that Eq. (1A) provided superior fits to all our data.

(Historically A. V. Hill introduced the basic form of Hill functions in 1910. It was widely used in a variety of applications in chemistry, biochemistry, physiology, and pharmacology, and the development of mathematical models of gene expression.[40])

We determined that $K_i$ has constant values of 395 and 4500 for all CdTe30 and CdTe300 combinations when covered by PMMA. The corresponding $H_i$ parameter values are 3.6 and 1.

These constant parameters indicate that the second term in the denominator of Eq.(1A) is only a function of energy, ignoring the effect of the overlying PMMA layer. This phenomenon is also visible in Figure 6(a,b), where all of the curves in the second section of the graph overlap. In contrast, the split curves in the first section of the graph necessitate different values for the parameters in the other term of the equation. Values of $K_a$ and $H_a$ for different thicknesses of PMMA added to CdTe are presented in Tables 1A and 2A.

Table 1A. Values of $K_a$ and $H_a$ parameters for different combinations of CdTe and PMMA

| PMMA Thickness (μ) | $K_a$ | $H_a$ |
|---|---|---|
| 0 | 0.5 ± 0.0 | 0.8 ± 0.2 |
| 1 | 19.2 ± 1.1 | 2.5 ± 0.5 |
| 10 | 59.5 ± 2.2 | 3.5 ± 0.4 |
| 50 | 144.0 ± 3.5 | 4.2 ± 0.4 |

Table 2A. Values of $K_i$ and $H_i$ parameters for different combinations of CdTe and Cu/Pb

| Metal Thickness (μ) | CdTe30+Cu | | CdTe30+Pb | | CdTe300+Pb | |
|---|---|---|---|---|---|---|
| | $K_i$ | $H_i$ | $K_i$ | $H_i$ | $K_i$ | $H_i$ |
| 0 | 390.0 ± 5.2 | 4.5 ± 0.1 | 370.0 ± 4.5 | 3.6 ± 0.1 | 2159.0 ± 42.0 | 2.0 ± 0.0 |
| 20 | 540.0 ± 10.8 | 3.6 ± 0.1 | 750.0 ± 8.0 | 3.8 ± 0.1 | 2885.0 ± 108.0 | 1.6 ± 0.0 |
| 50 | 670.0 ± 23.2 | 2.8 ± 0.1 | 1191.7 ± 17.7 | 3.9 ± 0.1 | 4233.0 ± 369.8 | 1.3 ± 0.1 |
| 100 | 716.6 ± 30.2 | 2.6 ± 0.1 | 1800.0 ± 0.0 | 3.8 ± 0.1 | 5506.0 ± 916.9 | 1.1 ± 0.1 |
| 200 | 738.0 ± 33.7 | 2.5 ± 0.1 | 2600.0 ± 187.0 | 3.1 ± 0.1 | 5867.0 ± 1092.6 | 1.1 ± 0.1 |

Trying to elucidate the physical meaning of the above fitting parameters we have plotted them in Figure 2A which suggested the following relation

$$\frac{K}{H} = a \cdot t^b \qquad \text{(Eq.2A)}$$

where parameters a and b are listed in Table 3A.

Table 3A. Values of parameters a, b in Eq. (2A)

| | a | b |
|---|---|---|
| PMMA | 7.0 ± 0.6 | 0.5 ± 0.0 |
| Cu | 77.3 ± 26.5 | 0.3 ± 0.1 |
| Pb | 20.3 ± 5.0 | 0.7 ± 0.1 |

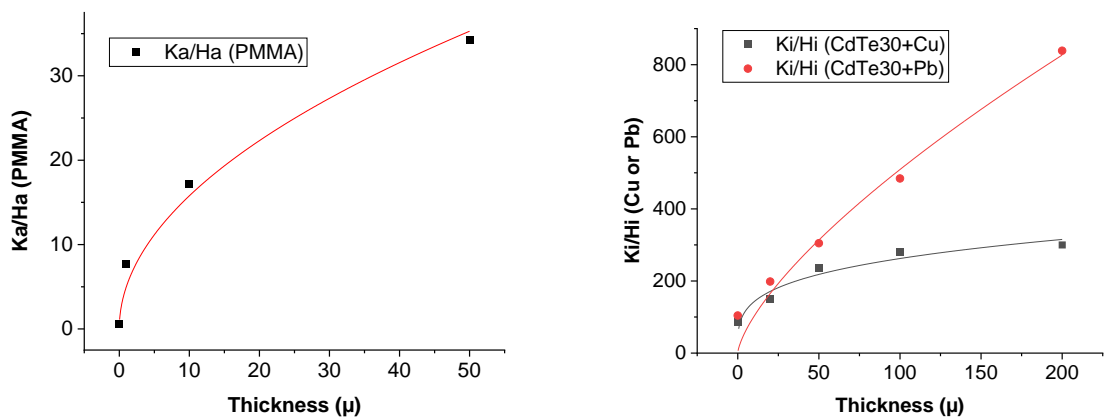

Figure 2A: Ratio between variable parameters in Eq.(1A) versus thickness of PMMA or metal back-reflectors; Left: $K_a/H_a$ for PMMA+CdTe structure, Right: $H_a/H_i$ for CdTe+Cu/Pb combinations

**Acknowledgments:**

We are grateful to V. G. Karpov for useful discussions of physical aspects pertaining to this work.

**Conflict of Interest:**
The authors have no relevant conflicts of interest to disclose

**Data availability statement:** Authors will share data upon request to the corresponding author.